\documentclass[twocolumn]{jpsj2} %% two-column layout
\def\runtitle{
Low-energy electrodynamics of heavy quasiparticles in ZrZn$_{2}$.
}
\def\runauthor{Shin-ichi \textsc{Kimura} \textit{et al}.}
\title{
Low-Energy Electrodynamics of Heavy Quasiparticles in ZrZn$_{2}$.
}
\author{
Shin-ichi \textsc{Kimura}$^{1,2,}$\thanks{E-mail address: kimura@ims.ac.jp}, 
Noriaki \textsc{Kimura}$^3$,
and Haruyoshi \textsc{Aoki}$^3$
}

\inst{
$^{1}$UVSOR Facility, Institute for Molecular Science, Okazaki 444-8585 \\
$^{2}$School of Physical Sciences, The Graduate University for Advanced Studies (SOKENDAI), Okazaki 444-8585 \\
$^{3}$Department of Physics, Tohoku University, Sendai 980-8578
}

\recdate
{
\today
}

\abst{
The temperature dependence of the optical conductivity spectrum of an itinerant ferromagnetic material, ZrZn$_{2}$, was obtained and the dynamical effective mass and the scattering rate were derived.
A renormalized Drude peak with a heavy effective mass developed at low temperatures.
The effective mass rapidly increased below $\hbar\omega$~=~10~meV and at 10~K it was observed to be 3 times heavier than that at 300~K.
The scattering rate also rapidly decreased below 10~meV at 10~K.
These results indicate the creation of heavy quasiparticles at low temperatures due to spin fluctuations.
}
\kword{ZrZn$_2$, optical conductivity, electronic structure, spin fluctuation}
\begin{document}
\maketitle
%
%%%%%%%%%%%%%%%%%%%%%%%%%%%%%%%%%%%%%%%%%%%%%%%%%%%%%%%%%%%%
\section{Introduction}
In recent years, some transition metal compounds that exhibit heavy-fermion-like behavior have attracted attention because of the associated interesting physical properties, such as superconductivity, that appear due to the strong electron correlation.~\cite{Nidda2003}
One of these materials, ZrZn$_2$, is known as an itinerant ferromagnetic material ($T_{\rm C}$ = 28~K) with a tiny magnetic moment ($\sim$~0.17~$\mu_{\rm B}$) in spite of consisting of only non-magnetic elements.~\cite{Mattias1958}
Extensive research into this phenomenon has revealed that this ferromagnetism originates from the spin fluctuations of the Zr $4d$ electrons.~\cite{Ueda1975}
Recently, this compound was discovered to process both superconductivity ($T_c$~=~0.29~K) and ferromagnetism.~\cite{Pfleiderer2001}
At low temperatures, heavy quasiparticles ($m^*\sim4.9m_0$) were observed through a de Haas-van Alphen (dHvA) experiment.~\cite{Yales2003}
The superconductivity of this compound is believed to originate from the strong electron correlation.
In other words, the spin fluctuations of the Zr $4d$ electrons are predicted to affect the heavy quasiparticles such that they exhibit superconductivity.
The coexistence of superconductivity and ferromagnetism is commonly observed in heavy fermion materials such as UGe$_2$ and URhGe.~\cite{Lohneyzen2002}
The same heavy fermion scenario in which the localized $f$ electrons hybridize with carriers is expected to influence the physical properties of ZrZn$_2$.

The superconductivity of $f$ electron systems appears around the quantum critical point (QCP) of the boundary between the local and itinerant regimes.~\cite{Coleman2005}
Near the QCP, the electrical resistivity is not proportional to $T^2$, which indicates a Fermi liquid.
In the case of ZrZn$_2$, the electrical resistivity near $T_{\rm C}$, which is proportional to $T^{5/3}$,~\cite{Yelland2005,Ogawa1976} suggests that ZrZn$_2$ is located near the QCP.~\cite{Uhlarz2004}
In a typical non-Fermi liquid heavy fermion material YbRh$_2$Si$_2$, non-Fermi liquid behavior has been revealed to be related to the strong spin fluctuation by electron spin resonance and nuclear spin resonance experiments.~\cite{Joerg2004, Ishida2002}
In the same temperature range, the scattering rate in the optical conductivity [$\sigma(\omega)$] spectrum of YbRh$_2$Si$_2$ is proportional to the photon energy, which is evidence of a non-Fermi liquid.~\cite{Kimura2006}
The photon-energy-dependent scattering rate and mass enhancement must appear in the $\sigma(\omega)$ spectrum if the electron correlation plays an important role in determining the physical properties at low temperatures.
According to the self-consistent renormalization (SCR) theory that explains itinerant ferromagnetism, the spin fluctuation effect appears not only in the magnetic properties but also in the transport properties, including the $\sigma(\omega)$.~\cite{Hasegawa1979}
Therefore, obtaining the $\sigma(\omega)$ spectrum provides important information in understanding the origin of heavy quasiparticles.

In this paper, the effect of the spin fluctuations on the $\sigma(\omega)$ of ZrZn$_2$ derived from the reflectivity spectra is reported.
Up to now, optical measurements of this material have been performed only in the narrow photon energy range of 0.6~--~3.8~eV.~\cite{Heide1984}
In this energy range, absorption due to interband transitions is mainly observed.
No information regarding heavy quasiparticles can be obtained in this range.
Therefore reflectivity [$R(\omega)$] spectra were then obtained over the wide energy range of 2~meV~--~30~eV at different temperatures from 10 to 300~K.
The temperature dependence of the $\sigma(\omega)$ spectrum was then obtained via the Kramers-Kronig analysis of the $R(\omega)$ spectra.
The effects of the spin fluctuations on the $\sigma(\omega)$ spectra and their relationship to heavy fermions are discussed.

%%%%%%%%%%%%%%%%%%%%%%%%%%%%%%
\section{Experimental and Band Calculation Methods}
The optical measurements were performed using a single ZrZn$_2$ crystal prepared by a slow cooling method~\cite{Schreurs1989} in a tungsten bomb with Zr($3N$) and Zn($6N$) as starting materials.~\cite{NKimura2004}
The near-normal incident optical $R(\omega)$ spectra of ZrZn$_2$ were acquired from well polished samples attained using 0.3~$\mu$m grain-size diamond wrapping film sheets.
Martin-Puplett and Michelson type rapid-scan Fourier spectrometers were used at photon energies ($\hbar \omega$) of 2~-~30~meV and 10~meV~-~1.5~eV, respectively, at sample temperatures between 10~-~300~K.
To obtain the absolute $R(\omega)$ spectra, reference spectra of the sample evaporated with gold {\it in situ} were then measured.
The $\sigma(\omega)$ spectra were derived from the Kramers-Kronig analysis (KKA) of the $R(\omega)$ spectrum in the wide energy range up to 30~eV measured at the beam line 7B of UVSOR-II, Institute for Molecular Science, Okazaki, Japan.~\cite{BL7B}
Since $R(\omega)$ above 1.2~eV has no significant temperature dependence, the $R(\omega)$ above 1.2~eV at 300~K was extrapolated to the experimental $R(\omega)$ obtained at lower temperatures below 1.2~eV.
In the energy ranges below 2~meV and above 30~eV, the spectra were extrapolated using the Hagen-Rubens function [$R(\omega)~=~1-(2\omega/\pi \sigma_{DC})^{1/2}$] and $R(\omega) \propto \omega^{-4}$, respectively.~\cite{Wooten}
After constructing $R(\omega)$ in the energy region from zero to infinity, a KKA was performed to obtain the $\sigma(\omega)$.

The band structure calculation was performed using the full potential linearized augmented plane wave plus the local orbital (LAPW + lo) method including spin orbit coupling implemented in the {\sc Wien2k} code.~\cite{WIEN2k}
ZrZn$_2$ forms a C15 cubic Laves crystal structure ($Fd-3m$, No.~227) with a lattice constant of 7.393~\AA.~\cite{Yelland2005}
The non-overlapping muffin-tin (MT) sphere radii values of 2.50 and 2.45 Bohr radii were used for the Zr and Zn atoms in ZrZn$_2$, respectively.
The value of $R_{MT}K_{max}$ (the smallest MT radius multiplied by the maximum $k$ value in the expansion of plane waves in the basis set), which determines the accuracy of the basis set used, was set to 7.0.
The total number of Brillouin zones was sampled with 4000~$k$ points.

%%%%%%%%%%%%%%%%%%%%%%%%%%%%%%%%%%%%%%%%%%%%%%%%%
\section{Results and Discussion}
\subsection{Temperature dependent reflectivity spectra}
%%%%%%%%%%%%%%  Reflectivity  %%%%%%%%%%%%%%%%%%%%
\begin{figure}[t]
\begin{center}
\includegraphics[width=0.4\textwidth]{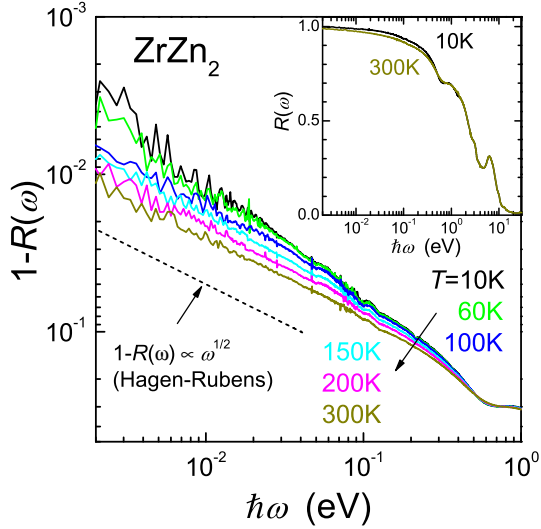}
\end{center}
\caption{
(Color online) The reflectivity spectra subtracted from unity [$1-R(\omega)$] in the photon energy range of 2~meV - 1~eV for different temperatures in the range of 10 -- 300~K.
The dashed line indicates the Hagen-Rubens function [$1-R(\omega)\propto\omega^{1/2}$].
Inset: $R(\omega)$ at 7 and 300~K in the photon energy range of 2~meV - 30~eV.
}
\label{Refl}
\end{figure}
%%%%%%%%%%%%%%%%%%%%%%%%%%%%%%%%%%%%%%%%%%%%%%%%%%
The temperature dependence of the reflectivity spectrum subtracted from unity [$1-R(\omega)$] is shown in Figure~\ref{Refl}.
The inset is the normal reflectivity spectra [$R(\omega)$].
The reflectivity spectra indicated metallic characteristics at all temperatures with two significant shoulder and peak structures at 1 and 5~eV.
These structures correspond to the interband transitions.

The high $R(\omega)$ intensity below 0.5~eV at all temperatures indicates that this material is a good metal.
At 300~K, the $R(\omega)$ curve at low energies obeys the Hagen-Rubens function [$1-R(\omega)\propto\omega^{1/2}$] as plotted in Figure~\ref{Refl}~\cite{DG} like a normal metal.
At lower temperatures, $R(\omega)$ increases with decreasing temperature below 0.4~eV, departing from the Hagen-Rubens relationship.
This indicates that the transport mechanism at lower temperatures changes from that at 300~K.
Note that a spectral change at $T_{\rm C}$ does not appear above the lowest accessible photon energy of 2~meV even though the exchange splitting is about 70~meV.~\cite{Yales2003,NKimura2005}
The discrepancy indicates that the density of states (DOS) relating to the ferromagnetic ordering is small compared with the high DOS at $E_{\rm F}$ as shown in Fig.~\ref{DOS}.
This is consistent with the small ordered moment of 0.17~$\mu_{\rm B}$ in the ferromagnetic state.

%%%%%%%%%%%%%%%%%%%%%%%%%%%%%%%%%%%%%%%%%%%%%%%%%
\subsection{Comparison with Band Calculation}
%
%%%%%%%%%%%%%%  DOS  %%%%%%%%%%%%%%%%%%%%
\begin{figure}[t]
\begin{center}
\includegraphics[width=0.4\textwidth]{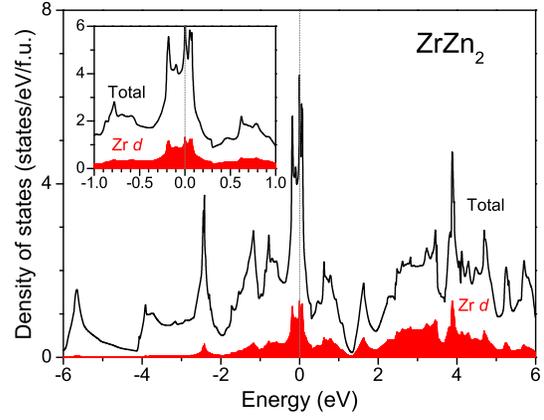}
\end{center}
\caption{
(Color online) The calculated density of states (DOS) of ZrZn$_2$.
The Zr~$d$ partial DOS is also plotted.
The inset is the detail in the energy range of -1.0 - 1.0~eV. 
}
\label{DOS}
\end{figure}
%%%%%%%%%%%%%%%%%%%%%%%%%%%%%%%%%%%%%%%%%%%%%%%%%%
The band structure of ZrZn$_2$~\cite{Goot1980} and the calculated $\sigma(\omega)$ spectrum have previously been compared with the experimental spectrum.~\cite{Heide1984}
However, the measured energy range was very narrow.
The $\sigma(\omega)$ spectrum was again calculated from the band structure calculation and compared with the experimental wide-range $\sigma(\omega)$ spectrum.
The calculated DOS shown in Figure~\ref{DOS} is similar to that previously reported.~\cite{Yales2003,Goot1980}
The important feature here is that the high DOS mainly consisting of the Zr~$4d$ states is located at $E_{\rm F}$ as shown in the inset of Figure~\ref{DOS}.

%%%%%%%%%%%%%%  CalcOC  %%%%%%%%%%%%%%%%%%%%
\begin{figure}[t]
\begin{center}
\includegraphics[width=0.40\textwidth]{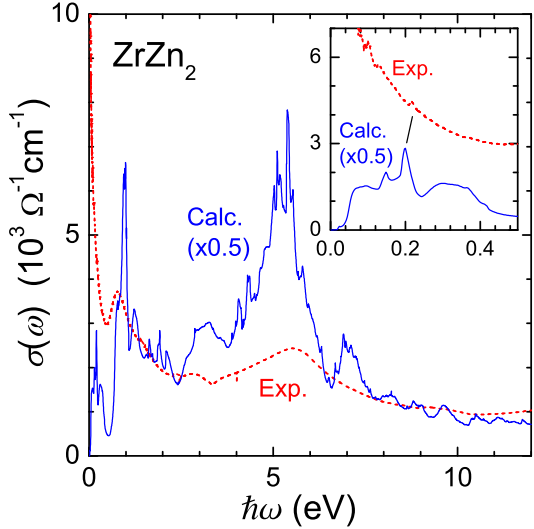}
\end{center}
\caption{
(Color online) Experimental optical conductivity [$\sigma(\omega)$] spectrum at $T$~=~300~K (dashed line) compared with the corresponding calculated one (solid line) for ZrZn$_2$.
The inset is an expansion of the low energy region, although the experimental spectrum was taken at 10~K.
}
\label{CalcOC}
\end{figure}
%%%%%%%%%%%%%%%%%%%%%%%%%%%%%%%%%%%%%%%%%%%%%%%%%%
To clarify the origin of the peaks of the experimental spectrum, the experimental $\sigma(\omega)$ spectra are compared with the theoretical ones obtained from the band structure calculation.
The $\sigma(\omega)$ spectra are derived from a function as follow in which direct interband transitions are assumed;~\cite{Ant04}
\[
\sigma(\omega) = \frac{\pi e^2}{m_0^2 \omega} \sum_{\vec{k}} \sum_{n n'} \frac{|\langle n' \vec{k}|\vec{e} \cdot \vec{p}|n \vec{k}\rangle |^{2}}{\omega - \omega_{n n'}(\vec{k})+i\Gamma} \times \frac{f(\epsilon_{n\vec{k}})-f(\epsilon_{n'\vec{k}})}{\omega_{n n'}(\vec{k})}
\]
Here, the $|n' \vec{k}\rangle$ and $|n \vec{k}\rangle$ states denote the unoccupied and occupied states, respectively, $\vec{e}$ and $\vec{p}$ are the polarization of light and the momentum of the electron, respectively, $f(\epsilon_{n\vec{k}})$ is the Fermi-Dirac distribution function, $\hbar\omega_{n n'}=\epsilon_{n\vec{k}}-\epsilon_{n'\vec{k}}$ is the energy difference of the unoccupied and occupied states and $\Gamma$ is the lifetime parameter.
In the calculation, $\Gamma$~=~1~meV was assumed.

The experimental $\sigma(\omega)$ spectrum of ZrZn$_2$ and the corresponding calculated spectrum are shown in Figure~\ref{CalcOC}.
In both the experimental and calculated spectra, the large peaks due to the interband transitions were located at 1 and 5~eV, as was seen in the $R(\omega)$ spectrum.
The peak at 0.2~eV in the experimental $\sigma(\omega)$ spectrum was also reproduced by the calculation, as shown in the inset of Figure~\ref{CalcOC}.
These results indicate that the calculated $\sigma(\omega)$ spectrum is consistent with the experimental one.
However, the intensity of the calculated $\sigma(\omega)$ spectrum is noted to be more than double the experimental one.
The same tendency has been observed in other materials,~\cite{Zerec2005} although the reason for this is not clear at present.

According to the calculated $\sigma(\omega)$ spectrum, the interband transition in the high DOS near $E_{\rm F}$ appears below $\hbar\omega$~=~0.4~eV, which energy corresponds to the width of the DOS.
The $R(\omega)$ spectrum shows temperature dependence below 0.4~eV (see Figure~\ref{Refl}) indicating a temperature dependent electronic structure of the large DOS near $E_{\rm F}$.

%%%%%%%%%%%%%%%%%%%%%%%%%%%%%%%%%%%%%%%%%%%%%%%%%
\subsection{Extended Drude Analysis}
%
%%%%%%%%%%%%%%  TdepOC  %%%%%%%%%%%%%%%%%%%%
\begin{figure}[t]
\begin{center}
\includegraphics[width=0.4\textwidth]{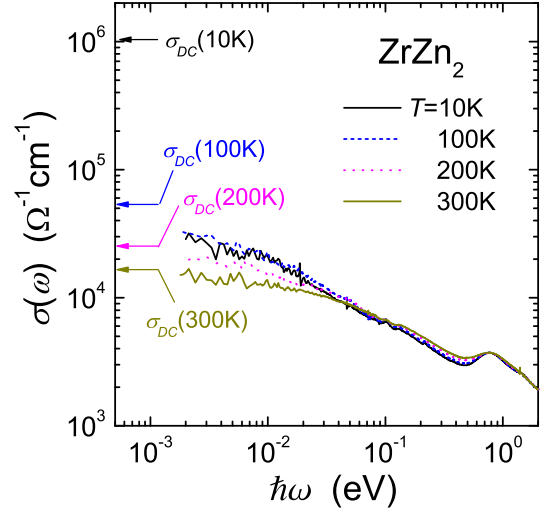}
\end{center}
\caption{
(Color online) Temperature dependent optical conductivity [$\sigma(\omega)$] spectra (lines) with corresponding direct current conductivity ($\sigma_{DC}$, arrows).
}
\label{TdepOC}
\end{figure}
%%%%%%%%%%%%%%%%%%%%%%%%%%%%%%%%%%%%%%%%%%%%%%%%%%
Figure~\ref{TdepOC} shows the $\sigma(\omega)$ spectra that are derived from the $R(\omega)$ spectra shown in Figure~\ref{Refl} via the KKA at typical temperatures and the corresponding direct current conductivity ($\sigma_{DC}$).
In Figure~\ref{TdepOC}, all of the $\sigma(\omega)$ spectra increase below $\hbar\omega$~=~0.5~eV with decreasing photon energy, indicating metallic Drude conductivity at all temperatures.
Above 100~K, the $\sigma(\omega)$ spectrum at the lowest accessible photon energy ($\hbar\omega$~=~2~meV) increases with decreasing temperature, and the extrapolation smoothly connects to the corresponding $\sigma_{DC}$ value.
On the other hand, the $\sigma(\omega)$ spectral intensity at 10~K is lower than that at 100~K despite the fact that the $R(\omega)$ spectrum at 10~K is higher than that at 100~K, as shown in Figure~\ref{Refl}.
In addition, the $\sigma_{DC}$ value is much higher than the $\sigma(\omega)$ intensity at 2~meV.
Both of these results and the spectral weight conservation rule indicate that a large but narrow Drude peak due to the existence of quasiparticles must appear below 2~meV as for other heavy fermion compounds.~\cite{Kimura2006,webb86,awa93,deg96}

%%%%%%%%%%%%%%  ExtDrude  %%%%%%%%%%%%%%%%%%%%
\begin{figure}[t]
\begin{center}
\includegraphics[width=0.4\textwidth]{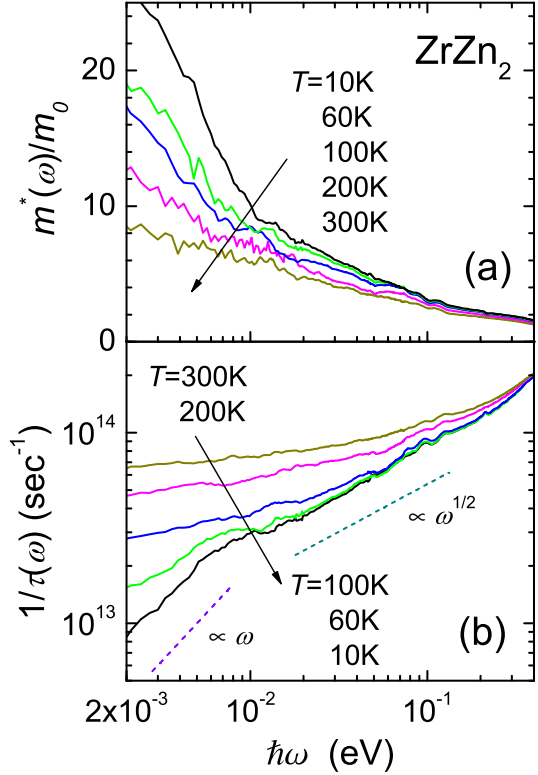}
\end{center}
\caption{
(Color online) Temperature dependence of (a) the effective mass relative to the free electron mass, $m^*(\omega)/m_0$, and (b) the scattering rate $1/\tau(\omega)$ as a function of the photon energy $\hbar\omega$ derived from the extended Drude analysis.
The spectra were smoothed by averaging in 50 data points ($\sim$~3~meV) for clarity.
The dashed lines in (b) emphasize $1/\tau(\omega) \propto \hbar\omega$ and $(\hbar\omega)^{1/2}$ behavior below and above 10~meV, respectively.
}
\label{ExtDrude}
\end{figure}
%%%%%%%%%%%%%%%%%%%%%%%%%%%%%%%%%%%%%%%%%%%%%%%%%%
To clarify the photon energy and temperature dependences of the quasiparticles, an {\it extended} Drude analysis in terms of the effective mass ($m^*$) and the scattering rate ($1/\tau$) was performed.
The coherent part of the underlying strong electron-electron correlations were treated in an {\it extended} Drude model by renormalized and frequency (photon energy) dependent $m^{*}(\omega)/m_{0}$ and $1/\tau(\omega)$;~\cite{webb86,awa93,kim94,deg96}
\[
\frac{m^{*}(\omega)}{m_{0}} = \frac{N_{eff} e^2}{m_0 \omega} \cdot Im\left(\frac{1}{\tilde{\sigma}(\omega)}\right),
\]
\[
\frac{1}{\tau(\omega)} = \frac{N_{eff} e^2}{m_0} \cdot Re\left(\frac{1}{\tilde{\sigma}(\omega)}\right).
\]
Here, $N_{eff}$ is the effective carrier density, $e$ the elementary charge, $m_{0}$ the electron rest mass and $\tilde{\sigma}(\omega)$ the complex optical conductivity derived from the KKA of the $R(\omega)$ spectrum.
$N_{eff}$ can be evaluated to be $8.0\times10^{21}$ cm$^{-3}$ through integration of the $\sigma(\omega)$ below the plasma edge of $\hbar\omega_p$~=~0.4~eV,
\[
N_{eff} = \frac{4 m_0}{h^2 e^2}\int^{\hbar\omega_p}_{0} \sigma(\hbar\omega) d\hbar\omega .
\]
The obtained $m^{*}(\omega)/m_{0}$ and $1/\tau(\omega)$ are plotted in Figure~\ref{ExtDrude}.

Even at 300~K, $m^*(\omega)/m_0$ is not constant.
The reason for this is that the interband transition appears at 50~meV~--~0.4~eV, as shown in the inset of Figure~\ref{CalcOC}.
This interband transition can not be subtracted from the experimental $\sigma(\omega)$ spectra because it creates a broad background.
Below 50~meV, not only $m^*(\omega)/m_0$ but also $1/\tau(\omega)$ are nearly constant as the $R(\omega)$ spectrum obeys the Hagen-Rubens function.
The effective mass of $8.0m_0$ at the lowest accessible energy of 2~meV at 300~K increases with decreasing temperature.
At 10~K, $m^*(\omega)$ becomes $25m_0$ at $\hbar\omega$~=~2~meV and $10m_0$ at 10~meV.
The mass enhancement factors ($1+\lambda=m^*(T)/m^*({\rm 300~K})$) at 2 and 10~meV are about 3 and 1.7, respectively.
In the previous dHvA experiment, the cyclotron mass enhancements of the quasiparticles were reported to be in the range of about 2~--~5.~\cite{Yales2003}
Therefore, the optically observed $m^*(\omega)/m_0$ at the experimentally accessible low energy limit is consistent with that at $E_{\rm F}$ detected by the dHvA.
$1/\tau(\omega)$ is almost constant below 50~meV at high temperatures, as noted previously.
At low temperatures, on the other hand, $1/\tau(\omega)$ decreases with decreasing photon energy.
Since $1/\tau(\omega)$ is the reciprocal relaxation time, the relaxation time increases with decreasing temperature and with decreasing photon energy.
This is direct evidence of the creation of quasiparticles at low temperatures.
Since the DC conductivity at low temperatures varies as $1-(T/T_a)^m$, with $m=5/3$ in the present case, the temperature dependence when including a photon energy dependence for the scattering rate can be written in the form,
\[
\frac{1}{\tau}=\frac{1}{\tau_a}\left[1-\left(\frac{T}{T_a}\right)^m-\left(\frac{\omega}{\omega_a}\right)^m\right],
\]
where $\omega_a$ and $T_a$ play the role of a cutoff frequency and temperature, respectively, and $\tau_a$ is a constant.~\cite{deg96,deg99}
At low temperatures, the electrical resistivity is proportional to $T^{5/3}$ near $T_{\rm C}$, which indicates non-Fermi liquid behavior, and to $T^2$ at very low temperature.
$1/\tau(\omega)$ in Figure~\ref{ExtDrude} at 10~K is proportional to the square root of the photon energy above 10~meV and is linear with photon energy below 10~meV.
Since the power law of the photon energy of $1/\tau(\omega)$ increases with decreasing photon energy, it must be proportional to $T^{5/3}$ below the lowest accessible photon energy of 2~meV near $T_{\rm C}$ according to the same power law in $T$ and $\omega$ as shown in the above function.
In the expectation from the SCR theory, $1/\tau(\omega)$ is almost constant below $\hbar\omega\sim\hbar/\tau(0)$ and the $\hbar\omega$-linear dependence should appear above $\sim\hbar/\omega(0)$.~\cite{Takahashi2007,Moriya1987}
The $\hbar\omega$-linear dependence in the experimental photon energy range is consistent with the theoretical expectation.

As discussed above, ZrZn$_2$ exhibits an increase in the effective mass and a rapid decrease in the scattering rate below 10~meV.
The photon energy of 10~meV can be regarded as the representative photon energy of the heavy quasiparticles, $\hbar\omega^*$.
Based on the $\sigma(\omega)$ of two-dimensional antiferromagnetic high-$T_c$ cuprates calculated using the SCR theory, the mass enhancement and the reduction of the scattering rate appear in the photon energy range below the photon energy equal to 0.1 times the typical temperature, $T_0$, of the spin fluctuation spectrum.~\cite{Moriya1987}
Therefore, $\hbar\omega^*$ should equal about $0.1 \times T_0$ in the two-dimensional antiferromagnetic case.
If the same logic can be attributed to a three-dimensional ferromagnetic material of ZrZn$_2$, $\hbar\omega^*$ must be about 11~meV (139~K) because of $T_0$~=~1390~K based on the SCR theory.~\cite{Takahashi2001}
This value is consistent with $\hbar\omega^*$ of ZrZn$_2$.
However, since it is not clear that the same logic can be attributed or not, the further theoretical investigation of the $\sigma(\omega)$ spectra of three-dimensional ferromagnetic spin fluctuation materials is needed.

A few other transition metal compounds with spin fluctuations, MnSi,~\cite{Mena2003} Sr$_2$RuO$_4$,~\cite{Katsufuji1996} and SrFe$_4$Sb$_{12}$,~\cite{Kimura2006-2} also show evidence of heavy quasiparticles in the $\sigma(\omega)$ spectra.
Since these materials are located near the QCP, the magnetic instability is believed to affect the transport properties.
This relates to the continuity of the Fermi surface at the QCP that is a topic in the heavy fermion physics of $f$-electron compounds.~\cite{Coleman2005}
If the SCR theory, which is based on band theory, can explain the physical properties not only of the transition metal compounds but also of the heavy fermion compounds at the QCP, the continuity of the Fermi surface at the QCP can be explained.~\cite{Coleman2005}
From this the physical properties including the above transition metals and heavy fermion compounds can be generally understood.

%%%%%%%%%%%%%%%%%%%%%%%%%%%%%%
\section{Conclusion}
In conclusion, the temperature dependence of the optical conductivity spectra of ZrZn$_2$ were ascertained in the wide photon energy range of 2~meV~--~30~eV and compared with band structure calculations.
The experimentally obtained spectra could be explained by the corresponding spectra derived from the band structure calculation.
At energies lower than 0.4~eV, a renormalized Drude peak with a heavy effective mass due to the spin fluctuations appeared at lower temperatures.
The effective mass rapidly increased with decreasing temperature and decreasing photon energy below the representative photon energy $\hbar\omega^*$ of 10~meV.
At 10~K, the effective mass reached a value about three times heavier than that at 300~K, which is consistent with the previous de Haas-van Alphen experiment.
The scattering rate, $1/\tau$, was also suppressed with decreasing temperature and decreasing photon energy.
$1/\tau$ at 2~meV was linear with the photon energy below 10~meV.
The linear-$\omega$ dependence is consistent with the SCR theoretical expectation.
This indicates that the optically observed heavy quasiparticles originate from the spin fluctuations.
To our knowledge, this is the first optical observation of heavy quasiparticles due to spin fluctuations in weak itinerant ferromagnetic materials.

%%%%%%%%%%%%%%%%%%%%%%%%%%%%%%
\section*{Acknowledgments}
We would like to thank Prof. Y. Takahashi for his fruitful discussion and suggestion of the SCR theory.
This work was a joint studies program of the Institute for Molecular Science (2006) and was partially supported by a Grant-in-Aids of Scientific Research (B) (No.~18340110) from MEXT of Japan.
%%%%%%%%%%%%%%%%%%%%%%%%%%%%%%

\end{document}